\documentclass[%
 reprint,
 amsmath,amssymb,
 aps,
]{revtex4-1}

\usepackage[utf8]{inputenc}
\usepackage{graphicx}
\usepackage{dcolumn}
\usepackage{bm}
\usepackage[usenames, dvipsnames]{color}
\usepackage{siunitx}
\usepackage{ulem}

\begin{document}

\preprint{APS/123-QED}

\title{Resonant Electro-Optic Frequency Comb}

\author{Alfredo Rueda}
\affiliation{Max Planck Institute for the Science of Light, Staudtstr. 2, 90158 Erlangen, Germany }%
\affiliation{Institute for Optics, Information and Photonics, University Erlangen-Nuernberg, Staudtstr. 7/B2, 91058 Erlangen, Germany}
\affiliation{SAOT, School in Advanced Optical Technologies, Paul-Gordan-Str. 6, 91052 Erlangen, Germany}%
\affiliation{Institute of Science and Technology Austria, am Campus 1, 3600 Klosterneuburg, Austria}%
\affiliation{The Dodd-Walls Centre for Photonic and Quantum Technologies, New Zealand}
\affiliation{Department of Physics, University of Otago, Dunedin, New Zealand}

\author{Florian Sedlmeir}
\affiliation{Max Planck Institute for the Science of Light, Staudtstr. 2, 90158 Erlangen, Germany }%
\affiliation{Institute for Optics, Information and Photonics, University Erlangen-Nuernberg, Staudtstr. 7/B2, 91058 Erlangen, Germany}

\author{Gerd Leuchs}
\affiliation{Max Planck Institute for the Science of Light, Staudtstr. 2, 90158 Erlangen, Germany }%
\affiliation{Institute for Optics, Information and Photonics, University Erlangen-Nuernberg, Staudtstr. 7/B2, 91058 Erlangen, Germany}
\affiliation{Institute of Applied Physics, Russian Academy of Sciences, 46 UL’yanov Street, 603950, Nizhny Novgorod, Russia}
\affiliation{The Dodd-Walls Centre for Photonic and Quantum Technologies, New Zealand}
\affiliation{Department of Physics, University of Otago, Dunedin, New Zealand}

\author{Madhuri Kumari}
\author{Harald G.\ L.\ Schwefel}
\email{harald.schwefel@otago.ac.nz}
\affiliation{The Dodd-Walls Centre for Photonic and Quantum Technologies, New Zealand}
\affiliation{Department of Physics, University of Otago, Dunedin, New Zealand}
%



\date{\today}

\begin{abstract}

\end{abstract}

\pacs{Valid PACS appear here}
\maketitle

{\bf High speed optical telecommunication is enabled by wavelength division multiplexing, whereby  hundreds of individually stabilized lasers encode the information within a single mode optical fiber. 
Higher bandwidths require higher a total optical power, but the power sent into the fiber is limited by optical non-linearities within the fiber and energy consumption of the light sources starts to become a significant cost factor \cite{kahn_communications_2017}. Optical frequency combs have been suggested to remedy this problem by generating multiple laser lines within a monolithic device, their current stability and coherence lets them operate only in small parameter ranges \cite{pfeifle_coherent_2014,ataie_ultrahigh_2015,marin-palomo_microresonator-based_2017}. Here we show that a broadband frequency comb realized through the electro-optic effect within a high quality whispering gallery mode resonator can operate at low microwave and optical powers. Contrary to the usual third order Kerr non-linear optical frequency combs we rely on the second order non-linear effect which is much more efficient. Our result uses a fixed microwave signal which is mixed with an optical pump signal to generate a coherent frequency comb with a precisely determined carrier separation. The resonant enhancement enables us to operate with microwave powers three orders of magnitude smaller than in commercially available devices. 
Such an implementation will be advantageous for next generation long distance telecommunication which relies on coherent emission and detection schemes to allow for operation with higher optical powers and at reduced cost \cite{temprana_overcoming_2015}.}

\begin{figure*}
\includegraphics[width=1.0\textwidth]{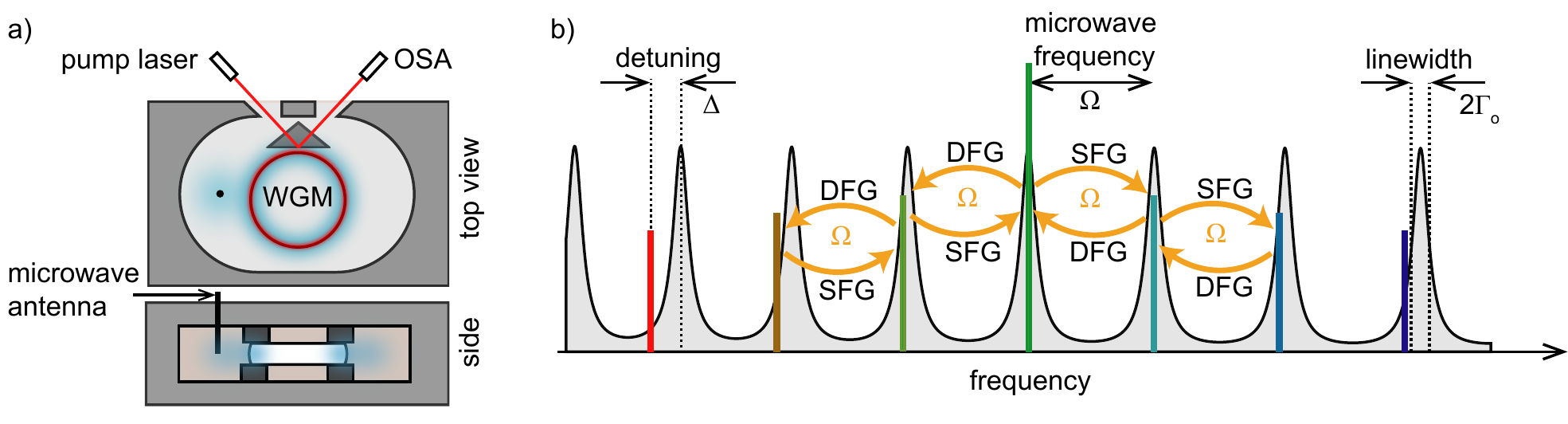}
\caption{\label{fig:theo}  \textbf{Principles of $\chi^{(2)}$ WGM-based frequency combs generation} a) Schematic of the setup for the creation of electro-optic $\chi^{(2)}$-frequency combs. The interacting fields are resonantly enhanced by an optical lithium niobate whispering gallery mode resonator (WGM) and an external 3D metallic microwave cavity. Light is coupled evanescently through frustrated total internal reflection into the WGM. The microwave {field} phase-modulates the light via the Pockels effect and a frequency comb is generated as illustrated schematically in b): The optical modes are separated by a nearly constant free spectral range (FSR) which approximately matches the frequency of the microwave tone $\Omega$ allowing for sum- and difference frequency generation (SFG, DFG) of sidebands. 
Due to the fixed microwave frequency, the comb lines are generated strictly equidistantly. This leads to a detuning of the comb lines from the optical modes, which are subject to dispersion, and eventually breaks the comb as illustrated in Figure \ref{fig:figure2}. 
 }  
\end{figure*}

The {data capacity of the internet} is expected to grow by a factor of two every year \cite{mitchell17}, but current optical techniques are not able to meet the rising demand {on} the bandwidth of the undersea fibre network \cite{Rev2018ofc}. Techniques such as space division multiplexing \cite{Kahn2017}, mode-division multiplexing \cite{Luo2014} and wavelength division multiplexing (WDM) \cite{temprana_overcoming_2015} in combination with time domain multiplexing (TDM) are being investigated {}to exploit the existing network to its full capacity. Current WDM systems employ an array of individually stabilized lasers, which are not phase locked to each other. For the next generation a major shift in the paradigm from multiple independent optical carriers to coherent optical frequency combs (OFCs) \cite{Rev2018ofc} combined with real time numerical calculation of the nonlinear pulse \cite{temprana_overcoming_2015} will be necessary. The advantage of OFCs is that they can be generated from a single laser, potentially reducing the overall energy consumption of the system considerably. Furthermore, depending on the method of comb generation, OFCs can feature high phase and frequency stability, and may also offer tunability of the comb line spacing.  
Of particular interest for future WDM applications is the intrinsic phase lock between all comb lines which allows the numerical counteracting of one of the main limitations, the nonlinear pulse distortion, for {accepting} higher optical powers within the telecommunication fibres \cite{temprana_overcoming_2015}. Commercial OFCs are currently based on femotosecond lasers \cite{Hansch2000,Liang2015}, however, over recent years combs generated in micro-resonators via the Kerr effect became more and more successful \cite{Kippenberg2014,suh_gigahertz-repetition-rate_2018,kippenberg_microresonator-based_2011,Liang2015}.

An alternative approach of comb generation is electro-optic modulation (EOM) \cite{Torres2014}. This scheme exploits a second-order non-linearity, whereby two continuous waves, e.g.\ one in the optical range (carrier) and one in the microwave range (modulating wave), are mixed within a non-linear non-centrosymmetric crystal. In the past, over-driving of conventional electro-optic modulators was explored for comb generation \cite{kovacich_short-pulse_2000,jiang_spectral_2007,wu_generation_2010,beha_electronic_2017}. These EOM based combs have unique advantages such as tunable central-frequency and comb-line spacing crucial for a range of applications ranging from telecommunication to spectroscopy. However, these combs suffer from inherent phase instabilities and excess multiplication of microwave phase noise \cite{beha_electronic_2017} which limits their usage in long-distance data transfer \cite{temprana_overcoming_2015}.

Here, we report the first experimental demonstration of efficient resonantly enhanced electro-optic comb generation. Such a fully coherent comb with intrinsic phase noise suppression allows to increase the information content and data rate in optical communication. The phase stability between the comb lines is inherently given due to the fixed interaction of two stable sources, high quality optical resonators filter the fundamental electro-optic noise \cite{beha_electronic_2017}.
Our experiment draws on an in-depth theoretical description of the resonantly enhanced non-linear interaction and we show in a first experiment the potential of this method by reporting on a coherent comb with 8.9\,GHz spacing with a span of almost 200 lines in the C-Band (1530\,nm -- 1565\,nm) with only 20\,dBm of microwave modulation power. The results clearly demonstrate a resurgence of EOM based comb generation and show the large potential of resonantly enhanced electro-optic modulation for the next generation of ultra dense WDM.
\begin{figure*}[t!]
\includegraphics[width=1.0\textwidth]{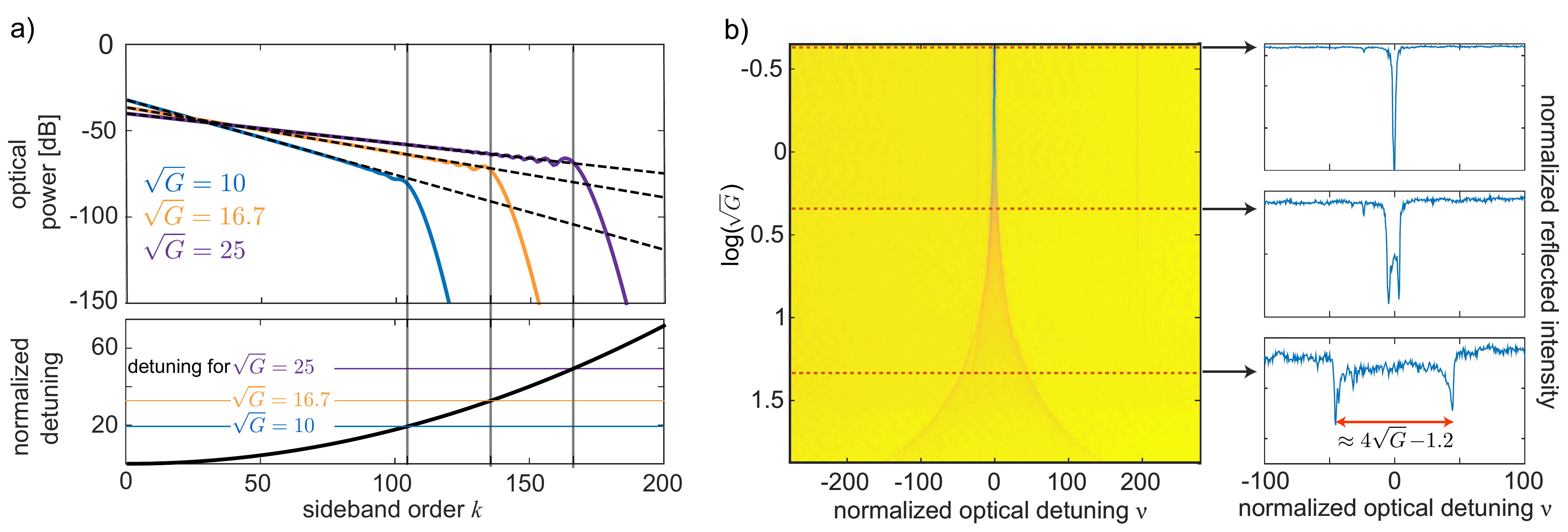} 
\caption{\label{fig:figure2}\textbf{Theoretical scaling of the sideband power and dispersion induced breakdown of the comb} a) shows the power of the sidebands as a function of the sideband order for different modulation strengths ($G$). The dashed lines represent the analytic solution given by equation (\ref{eq:fourier}) and the straight colored lines are numeric solutions which include the optical dispersion of lithium niobate (see Supplement). The dispersion leads to a detuning of the equidistantly generated comb lines from the optical resonance frequencies as shown in the bottom part. When a certain detuning of the comb lines from the optical resonance frequency is reached, the comb breaks down drastically. The $G$ dependence of the breakdown can be understood by the nonlinear induced broadening of the modes shown in b): in the yellow contour plot, we show the measured mode splitting as a function of $G$, which is proportional to the microwave power sent to the system. }
\end{figure*}

The scheme we use for comb generation is based on the Pockels effect: an electric field applied to a non-centrosymmetric crystal such as lithium niobate, results in a change of the optical refractive index of the material directly proportional to the applied voltage. As a consequence, light passing through the crystal encounters a varying optical path length and is thus phase modulated with the frequency of the applied voltage. This phase modulation generates sidebands separated from the optical carrier frequency by multiples of the modulation frequency and forms a comb. One can observe this behavior already in a standard EOM \cite{kourogi_wide-span_1993,ye_optical_2003}, but the magnitude of the sidebands usually decay quickly with distance from the carrier. The length of the comb is determined by the modulation index which scales with the amplitude of the applied electric field \cite{pozar_microwave_2011}.
To boost the electric field and thus the efficiency of the process and generate a wide spanning comb, we embed a high quality optical whispering gallery mode resonator into a high quality microwave resonator. 

Another, equivalent way to view the process is sum and difference frequency generation: the energy of the microwave photons is either added to or subtracted from the carrier photon yielding new frequencies. These sidebands are themselves interacting with microwave photons and give rise to new frequencies again by the same process.  
For an efficient comb generation the sidebands have to match an optical resonance, as a consequence, the microwave resonator needs to be designed such that its resonance frequency coincides with the free spectral range (FSR) of the optical resonator. Mathematically the system is described by an infinite but simple set of linear equations each representing an optical mode participating in the nonlinear process. It can be solved analytically under the condition of a constant FSR (see supplement and \cite{ilchenko_whispering-gallery-mode_2003}), and for the time dependent optical field amplitudes in reflection of the resonator, we find
\begin{eqnarray}
\frac{A_{\text{out}}(t)}{A_\text{in}(t)}&=&\frac{ 2\gamma/\Gamma-1 -i(\Delta +2\sqrt{G}\cos(\Omega t))}{1+i(\Delta+2\sqrt{G}\cos(\Omega t))}.\label{eq:allmodes}
\end{eqnarray}
Here $\Gamma = \gamma + \gamma'$ is the half linewidth of the optical modes with $\gamma$ and $\gamma'$ being the optical coupling rate and intrinsic field loss rate, respectively. The detuning of the optical carrier frequency $\omega_0-\omega$ from its resonance frequency $\omega_0$ is normalized to half of the optical linewidth $\Delta=(\omega_0-\omega)/\Gamma$ and $\Omega$ is the microwave frequency which we assumed to be on resonance. The figure of merit is $G = {n_\Omega} g^2 / |\Gamma|^2$, where $n_\Omega$ is the number of microwave photons in the resonator and $g$ is the single photon coupling rate which is a function of the electro-optic coefficient and the overlap between the optical and the microwave modes. $G$ is often referred to as electro-optic cooperativity \cite{tsang_cavity_2010} and represents the ratio between the nonlinear photon conversion rate and photon decay rate and is hence a measure for the strength of the nonlinear interaction.
It is interesting to compare the modulation of a resonant EOM with a non-resonant EOM which is simply described by $A_\text{out}(t) = e^{i \xi \cos(\Omega t)} A_\text{in}(t)$ with the modulation index $\xi \approx 4\sqrt{G}$. The non-resonant EOM does not contain any amplitude modulation, while the resonant EOM, on the other hand, does. Only when the resonant EOM is strongly over-coupled ($\gamma \gg \gamma'$) and for very small nonlinearities $G\approx 0$ are the two systems are comparable. At critical coupling ($\gamma = \gamma'$) and increasing $G$, the resonant EOM  starts generating pulses with a repetition rate of $1/2\tau$ where $\tau$ is the roundtrip time of the optical cavity. We show in the supplement that the pulse width $\Delta\tau$ can be approximated as $\Delta\tau \approx \tau/(2\pi \sqrt{G})$.

The power of the comb lines can also be analytically described as
\begin{equation}
\frac{P(\omega \pm k\Omega)}{P_\text{in}}=\frac{4\gamma^2}{\Gamma^2}\left|\frac{e^{-\beta(G)}}{\sqrt{G}+\sqrt{G}e^{-2\beta(G)}}\right|^2 e^{-2|k|\beta(G)}\mbox{ for  } k\ne0\label{eq:fourier},
\end{equation}
where $k$ represents the order of the sideband, $P^\text{in}$ is the input optical power and $2\beta(G)$ is the decay constant and can be approximated to $1/\sqrt{G}$ in the limit of strong nonlinear interaction $4G\gg1$ (see supplement). The resonant comb scales strictly exponential, which is another difference to the non-resonant comb, where the comb line power is described by Bessel functions \cite{hobbs2011building}. In Figure \ref{fig:figure2}a we plot Eq.~(\ref{eq:fourier}) for different modulation strengths {as dashed lines} on a logarithmic scale.  It is apparent that the slope of the exponential decay decreases with increasing $G$ leading to the generation of longer combs.
\begin{figure*}
\includegraphics[width=1.0\textwidth]{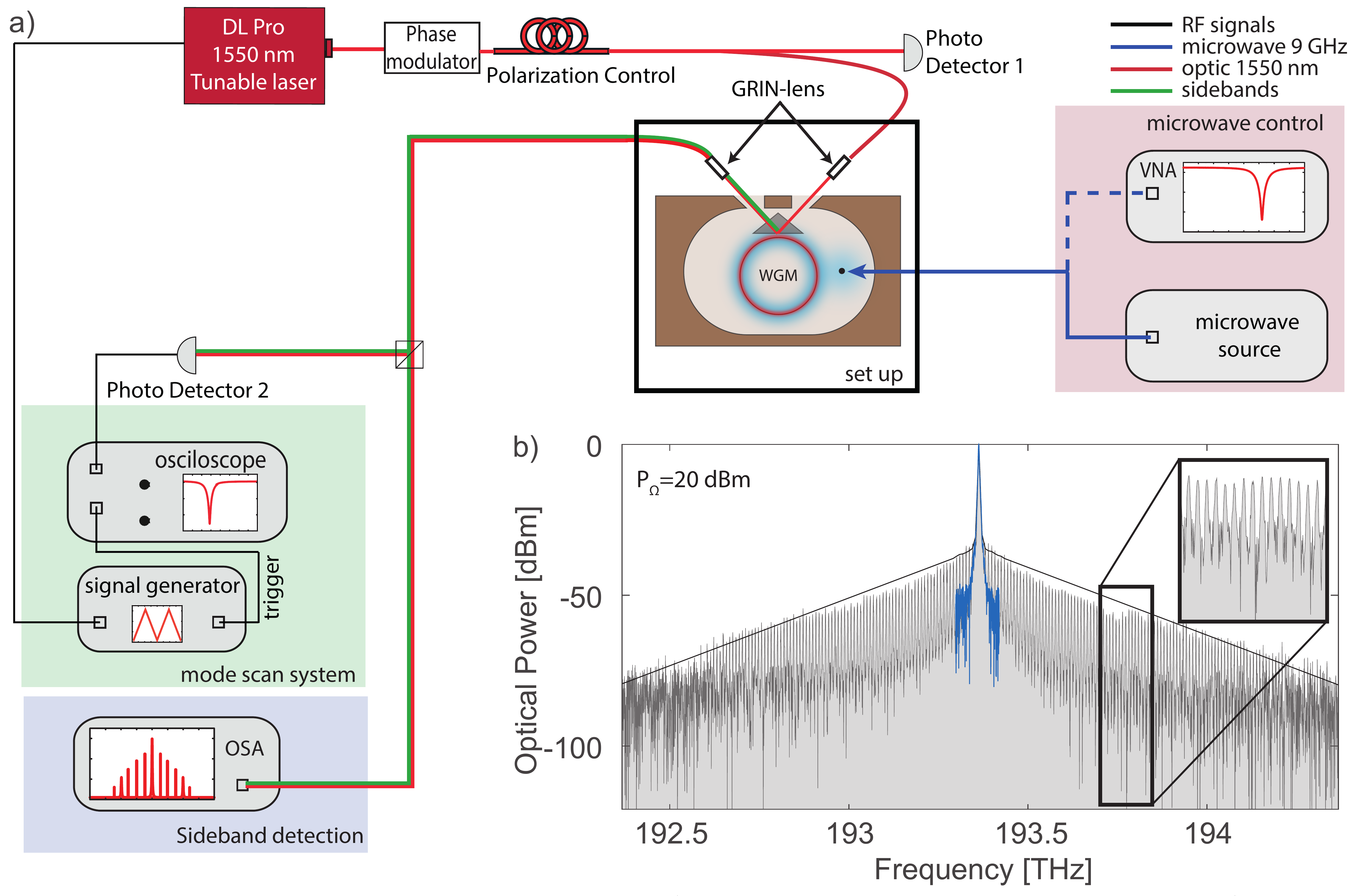}  
\caption{\label{fig:comb}\textbf{Experimental realization } a) Measurement Network. A monochromatic coherent laser source in the telecommunication domain goes through a phase modulator, used for measuring the optical free spectral range and linewidth of the modes.  The polarization controller sets the GRIN-lens output beam to TE polarization. 99\% of the light is coupled into the resonator, while 1\% of the light is used for power calibration (photo detector~1). The reflected optical carrier is sent to PD2 to characterize the optical spectrum of the resonator. The comb lines, also out-coupled through the silicon prism, are detected on an optical spectrum analyzer (OSA). The microwave tone was sent through a cable to the pin coupler inside the cavity. This transmission line was also used to characterize the microwave mode with a vector network analyzer (VNA). b) Spectrum of the frequency comb with 20\,dBm microwave power at 8.9\,GHz spanning over 180 comb lines resulting in a width of about 1.6\,THz. In blue we shown the spectrum of the pump laser without the microwave tone.}
\end{figure*}
To derive the previous analytic solution for the system, we neglected 
optical dispersion, which is intrinsic due to geometric and material dispersion. As a consequence, the optical modes are not exactly equally spaced and the generated comb lines become increasingly detuned from the optical resonance frequencies with distance from the carrier mode (compare Figure \ref{fig:theo}b). This eventually breaks the comb generation as illustrated in Figure \ref{fig:figure2}a. The straight lines show the sideband power as function of the sideband order obtained from numerically solving the rate equations including optical dispersion of lithium niobate \cite{zelmon_infrared_1997} (see supplement for details). Close to the pump frequency the power decreases exponentially as expected from equation (\ref{eq:fourier}) till a certain $G$ dependent sideband order where it drastically drops. This can be intuitively understood from Figure \ref{fig:figure2}b: with increasing microwave power, the optical modes broaden and show a mode splitting by $(4\sqrt{G} - 1.2) \Gamma$. As a consequence, the dispersion induced detuning of the optical eigenfrequencies from the generated sidebands impacts the comb generation considerably not already at the half-width of a cold cavity mode ($\Gamma$) but at $2\sqrt{G} - 0.6$. Figure \ref{fig:figure2}b shows that this intuitive picture is accurate. As soon as the detuning, shown in Figure \ref{fig:theo}b, equals the nonlinear broadening, the comb starts to break  down. Nevertheless, with reasonable parameters we expect a few hundred lines as shown in Figure \ref{fig:figure2}a.


To experimentally demonstrate our highly efficient scheme, we embedded a high quality resonator machined from single crystalline lithium niobate into a microwave cavity made from copper.
The optical resonator is a so called whispering gallery mode resonator manufactured from lithium niobate: a convex shaped disk which guides light via total internal reflection along its inner surface interfering with itself after each roundtrip \cite{reviewdmitry}. Since lithium niobate is quite transparent, the light can do many roundtrips before being absorbed which boosts the optical intensity by orders of magnitude facilitating efficient nonlinear interactions \cite{strekalov_microwave_2009,Rueda:16,botello_sensitivity_2018}. 
The microwave resonator is a 3D copper cavity enclosing the optical resonator designed such that the resonantly enhanced microwave field has maximum overlap with the optical whispering gallery mode. 
The cavity contains two protruded copper rings facing each other which clamp the optical resonator when the cavity is closed. This ensures that the microwave field is focused on the rim of the optical resonator where the optical modes are located maximizing the overlap between the two vastly different frequencies.
Two holes are used to couple optical light into and out from the cavity. Apart from that the copper cavity is closed to prevent the microwave mode from radiating into the far field which would decrease its {quality} $Q$ factor. Inside the cavity, we use a silicon prism placed close to the optical resonator to couple the light into the lithium niobate disk. 
The WGM resonator has a radius of $2.45\,$mm and a thickness of 0.4\,mm resulting in a free spectral range of about $8.9\,$GHz at the used pump frequency $\omega_0 = 2\pi \times 193.5\,$THz. Based on numerical simulations of the microwave mode ({see Methods}), we designed the copper cavity such that the microwave resonance matches approximately the optical free spectral range and added a fine tuning mechanism to compensate for small fabrication tolerances.
Our system has a single photon coupling rate of $g = 2\pi \times 7.43$ Hz, the optical and microwave quality factors are $Q_o =1.4\times10^8$ and $Q_\Omega =241.4 $ respectively. The optical resonator is critically coupled resulting in $\gamma = \gamma' = 2\pi \times 0.35\,$MHz while the microwave resonator is undercoupled with $\gamma_\Omega = 2\pi \times 3.6\,$MHz and $\gamma_\Omega' = 2\pi \times 16.20\,$MHz.
For convenient comparison with canonical electro-optic phase-modulators, one can estimate the $\pi-$voltage from these parameters to be $V_\pi = 260$ mV (see supplement). Typical non-resonant modulators have values of a few volts showing the high efficiency of our system. This is traded, of course, by having only a few MHz bandwidth compared to several GHz bandwidth of non-resonant systems.

To observe the comb, we couple a  317 $\mu$W of optical power into a fundamental whispering gallery mode and modulate it with $20\,$dBm of microwave power ($\sqrt{G} \approx 25$). The light emitted from the cavity is coupled to an optical fiber and measured with an optical spectrum analyzer. Figure \ref{fig:comb}b shows the generated comb to have a span of about $11\,$nm corresponding to 1.6\,THz and 90 visible comb lines in each direction. The comb is, as expected, symmetric and decays approximately exponential. According to our theory, we expect a dispersion induced breakdown of the comb at a span of 160 lines which we cannot observe due to the noise floor of the OSA.

In summary, we have demonstrated that multi-resonant electro-optic modulation can lead to the formation of a broadband frequency comb at very low electrical power consumption. We present a complete analytic solution for a dispersionless system describing the comb formation thoroughly, including phase and amplitude of the generated comb lines and its temporal behavior in steady state. Our numerical simulations show that even if considering optical dispersion, the comb can span hundreds of lines due to nonlinear line-broadening of the optical modes before breaking down.
Since two stabilized sources are used to generate such an electro-optic comb via a second order non-linear effect, the resulting comb lines have a fixed and pre-determined phase relation to each other. We believe that the combination of a compact high quality resonant enhancement with a careful microwave field engineering and a fast way to calculate the nonlinear pulse propagation in real time can become a key concept for power-efficient optical interconnects and extend the range of long distance interconnects due to the inherent phase relation of all the comb lines.

\section*{Acknowledgments}

\noindent This work was supported by the Marsden Fund of the Royal Society of New Zealand, the Julius von Haast Fellowship of the Royal Society of New Zealand, and the Max Planck Institute for the Science of Light, Erlangen, Germany.

\section*{Author Contributions}

\noindent A.R. and F.S. performed all the experiments and developed the theory. A.R. performed the theoretical and numerical modeling. H.G.L.S. proposed the experiment. A.R., F.S., M.K., G.L., and H.G.L.S. wrote the manuscript. All authors contributed to discussing and interpreting the results.

\section*{Additional information}

\noindent Correspondence and requests for materials should be addressed to H.G.L.S.

\section*{Competing financial interests}

\noindent The authors declare no competing financial interests.



\clearpage
\subsection{Methods}
\begin{figure*}
\includegraphics[width=0.6\textwidth]{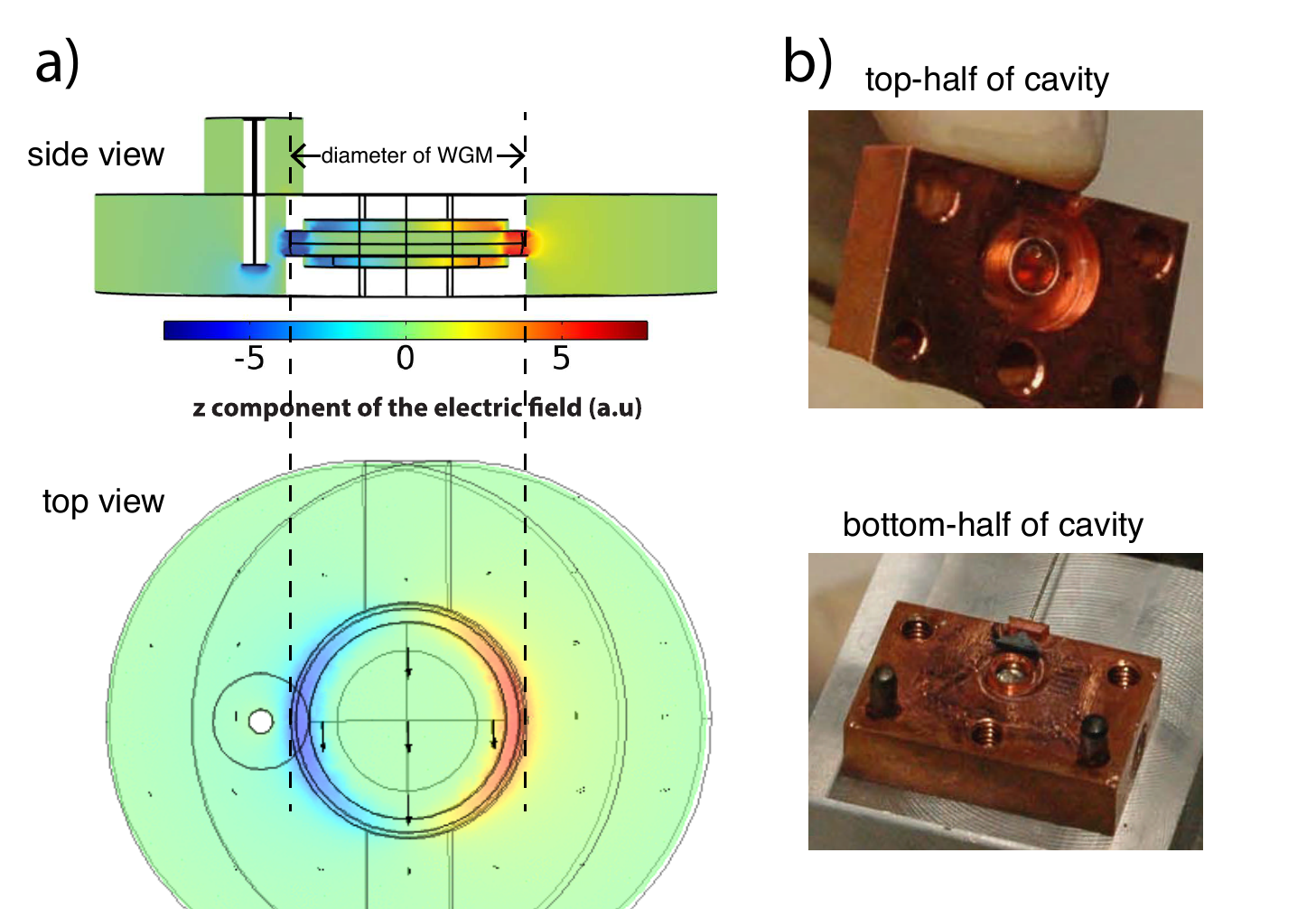}
\caption{\label{fig:methods}{ \textbf{Experimental resalisation } a) COMSOL simulation and b) top, and bottom-half of the copper cavity. The lithium niobate WGM is mounted in the top-half. }}
\end{figure*}

\textbf{Theory} 
Theoretically we can describe the system of a strong microwave field coupled nonlinearly to two optical fields by their total energy, its Hamiltonian. {Assuming no depletion of the microwave field (due to the presence of both sum- and difference generation) the nonlinear interaction Hamiltonian is given by (see supplement):}
\begin{equation}
\hat{H}_\text{int}=\hbar |\alpha| g (\hat{a}_1\hat{a}_2^\dagger +\hat{a}_1^\dagger\hat{a}_2). \label{eqMethods:hamscat}
\end{equation}
This is also known as the beam splitter Hamiltonian where $|\alpha|^2$ is the number of microwave photons in the cavity, $\hat{a}_1$ and $\hat{a}_2^\dagger$ are the annihilation and the creation operators of the two optical fields and can represent sum- as well as difference frequency generation. This linearisation of the Hamiltonian is also known as a zero-photon process characterized by photons scattering from one mode into another without changing the total number of photons \cite{Law}. 
The single photon coupling rate $g$ is defined as
\begin{equation}
g=2\epsilon_0 \chi^{(2)}  \sqrt{\frac{\hbar \omega_0\omega_s\omega_\Omega}{8\epsilon_0\epsilon_s\epsilon_\Omega V_0V_sV_\Omega}} \int dV \psi_\Omega \psi^\dagger_s\psi_0,\label{eqMethods:defgg}
\end{equation}
where $\chi^{(2)}$ is the second order nonlinear susceptibility, $\epsilon_0$ the vacuum permittivity, and $\omega_i,\epsilon_i,V_i, \psi_i$ are the frequency, relative permittivity, mode volume and normalised electric field distribution of the optical pump, sideband and microwave tone indicated by $i \in\{0,s,\Omega\}$, respectively. The integral over the fields is only nonzero if the so called phase-matching relation between the individual fields is fulfilled which leads to the relation $m_s=m_0+ m_\Omega$, where $m$ is the angular momentum number for the sideband, pump and microwave field, respectively (see supplement).
In a system with multiple modes $\omega_k$, equidistantly separated from the optical pump field frequency $\omega_0$, a strong microwave field {leads} to {a} cascade of the sum- and difference frequency generation. Assuming the same nonlinear coupling constant $g$ between the optical and microwave fields and noting that due to the phase-matching only {spectrally adjacent} modes can interact, the Hamiltonian can be written as:
 \begin{equation}
\hat{H}_\text{int}=\hbar |\alpha| g \sum\limits_{k=-N}^{N-1}(\hat{a}_k\hat{a}_{k+1}^\dagger +\hat{a}_k^\dagger\hat{a}_{k+1}), \label{eqMethods:hamsqu}
\end{equation}
where $2N+1$ is the number of optical modes involved in the system and $\hat{a}_k$ denotes the operator of the $k$ participating optical modes.
To describe their dynamics under the presence of a coherent microwave drive, we find the equations of motion using the Heisenberg picture to be:
\begin{equation}
\dot{\hat{a}}_k= \frac{i}{\hbar}[\hat{H}_\text{int},\hat{a}_k]=-i\sqrt{n_\Omega}g(\hat{a}_{k-1}+\hat{a}_{k+1}),\label{eqMethods:eqsum}
\end{equation}
where $n_\Omega$ is the number of microwave photons in the system. We introduce coupling and loss channels and solve these equations for the steady state in the rotating wave approximation and classical limit. Taking into account that we need to out couple the fields from the resonator we get equations (\ref{eq:allmodes}) and (\ref{eq:fourier}) in the main text.

\textbf{Numerical cavity design} 
{The microwave cavity was designed with the help of numerical simulations using COMSOL. The goal was to maximize the overlap between microwave and optical fields while choosing the system geometry such that the optical free spectral range coincides with the $m_\Omega$ = 1 microwave mode with a frequency around $10\,$GHz. In compliance with these boundaries, we optimized the microwave coupling and allowed for an optical coupling port to the WGM resonator.}
{We found a solution for efficient} type I phase matching ($\text{TE}+\text{TE} \rightarrow \text{TE}$) {for which the electro-optic coefficient of lithium niobate is largest}.  We note that this hybrid system is very similar to our previous work on coherent microwave up-conversion, showing the versatility of this approach \cite{Rueda:16}.




\textbf{Experiment} We fabricate a lithium niobate WGM resonator ($n_\text{o}=2.211$, $n_\text{eo}=2.138$, for the ordinary and extra-ordinary polarisation) with $R=2.45$\,mm and minor radius $r=1.5$\,mm, and height $h=0.4$\,mm, mounted within the copper cavity shown in Fig.~\ref{fig:methods} a,b) (see also supplement). As the optical pump we use a narrow-band tunable laser ({Toptica} DL pro, $\lambda \approx 1550$\,nm, linewidth of 100\,kHz) which goes through a polarization controller. {The pump} is split 99:1, with the weak {port} sent to the detector to keep {track} of the {pump power} (see Fig.~\ref{fig:comb} c). The other arm goes to the graded index lens for WGM-prism coupling to a transverse {(TE)} electric mode. We use a silicon prism ($n_\text{Si}=3.47$) mounted within the copper cavity. Two holes in the copper cavity allow the optical pump light to enter and the light reflected and the emitted sideband from the WGM resonator to leave the cavity. The {signal and reflected pump are coupled to a} fiber and sent to {an} InGaAs photodetector PDA10CS {(Thorlabs)} and its output signal to the oscilloscope. We can use a signal generator to sweep the lasers frequency and scan over the optical modes of the cavity, measure the optical FSR and optical loaded $Q$ using sideband spectroscopy \cite{Li:12}. 
The microwave signal {generated by an SMR20, Rohde \& Schwarz,} is coupled via a coaxial pin coupler introduced close to the WGM resonator inside the cavity. The same transmission line was used for reflection measurements with a vector network analyzer ENA E5072A, Keysight to obtain central frequency, coupling and loss rate of the microwave mode.  A metallic tuning screw is used to perturb the microwave field for fine adjustment of its resonance frequency (see Fig.~\ref{fig:figure2} a). The whole setup is thermally stabilized at $30^\circ$ to the mK level with a proportional-integral-derivative (PID) controller TC200 in combination with temperature sensor (AD590) and a thermoelectric element attached on the outer side of the closed copper cavity. 
Once the laser is set to the mode, we send the microwave tone and create the sidebands (see Fig.~\ref{fig:comb} d). These signals are outcoupled through the prism and coupled to the fibers. {The sidebands, together with the reflected optical pump are measured with an} OSA (YOKOGAWA AQ6370C).
\bibliography{firstbib}
\bibliographystyle{ieeetr}

\onecolumngrid
\clearpage
\section{Supplementary Material}
The structure of this supplementary material is as follows. First we describe optical whispering gallery mode (WGM) resonators and discuss the modal distribution. Then we introduce the mathematical formalism to describe a second order nonlinear interaction in the Hamiltonian description and consider the phase matching. We derive and solve the suitable rate equations in the Heisenberg picture. Finally we compare the efficiency of the process to the $\pi$-Voltage and discuss the experimental microwave coupling.

\subsection{Optical Whispering Gallery Modes \& Dispersion}
A whispering gallery mode resonator usually consists of a convex shaped dielectric material with refractive index $n$ higher than its surrounding \cite{reviewdmitry}. At an angle beyond the critical angle the light undergoes total internal reflection {and is guided on a circular path along the rim without refractive loss}. Light can be coupled into the WGM through an evanescent field coupler such as a prism. In a physical resonator loss is always present and we distinguish between intrinsic loss {rate} $\gamma'$ and coupling loss {rate} $\gamma$ due to the interaction of the prism. If the resonator is critically coupled the coupling loss equals the intrinsic loss $\gamma'=\gamma$ and {the photon number in the resonator is maximized}. The total loss rate is $\Gamma=\gamma+\gamma'$. In a prism coupled WGM system we have control over the coupling rate $\gamma$ by adjusting the distance of the prism from the resonator. Through careful adjustment, we can over-couple ($\gamma > \gamma'$) and under-couple ($\gamma < \gamma'$) the resonator.
The total loss rate is related to the photon lifetime in the resonator as $\tau_p=1/2\Gamma$ considering that the rates are defined for the field. The quality factor is $Q=\omega\tau_p$, where $\omega$ is the frequency of the resonance. The free spectral range corresponds to the inverse of the roundtrip time
\begin{equation}
\text{FSR}=1/{\tau}={c}/{(2\pi n {R}})\label{eq:FSR},
\end{equation} where $R$ and $c$ are the radius and the speed of light, respectively.

In a linear resonator the resonant frequency simply follows from the argument that an integer number of wavelengths needs to fit into the optical path length $\lambda_0=2\pi nR/m$  or equivalently $\omega=2\pi \cdot m \cdot\text{FSR}$ and $nk_0= m/ R$, with the wavenumber in free space $k_0=\omega/c$. For very large WGM resonators this might still be a valid expression. It does not however take the actual geometric dispersion seen in a three dimensional WGM into account. 
 In order to find better expressions for WGM resonators we need to consider the Helmholtz equation, with the appropriate boundary conditions given by the geometry. Here we choose a local toroidal coordinates system following Ref.~\cite{Breunig:13,Foreman:16}. In such coordinates the scalar field inside of the resonator and outside of the resonator can be written as:

\begin{equation}
\textbf{E}^\eta_{q,p,m}(\textbf{r}(\rho,\varphi,\theta))\approx  
\begin{cases} 
E_0 \exp(-\frac{\theta^2}{2\theta^2_m})H_p(\frac{\theta}{\theta_m})\text{Ai}[f^\eta_{m,q}(\rho)] e^{im\varphi}&\mbox{if } \rho<\rho'  \\
E'_\eta \exp(-\frac{\theta^2}{2\theta^2_m})H_p(\frac{\theta}{\theta_m})\exp(-\kappa(\rho-\rho')) e^{im\varphi}& \mbox{if } \rho>\rho' ,\label{WGMeq}
\end{cases}
\end{equation}
where $r(\rho,\varphi,\theta)$ is the position relative to the toroidal coordinates origin $\rho$ and $\theta$ stand for the distance and polar angle relative to the center of the curvature of the rim  { and $\rho'$ is the lateral radius of the WGM resonator}. $\eta$ is the polarization index parallel (TE) and orthogonal (TM) to the symmetry axis, $\kappa_\eta=k_0\sqrt{n^2_{\eta}-1}$ the evanescent decay constant. $q=\{1,2\dots\}$ is the radial number which corresponds to the number of maxima of the electric field intensity along the $\hat{r}$ axis. $p=\{0,1\dots\}$ stands the polar number and $m=\{1,2\dots\}$ is the azimuthal number representing the number of the electric field amplitude maxima along $\hat{\varphi}$. $H_p$ are the Hermite polynomials of degree $p$ which determine the lateral mode profile. The Airy function determines the radial distribution of the electric field in the resonator. $E_0$ is a normalization constant and $E'_\eta$ is the evanescence field amplitude given by the boundary conditions according to each polarization. The abbreviations in the formula are defined as 
\begin{subequations}
 \begin{eqnarray}
f^\eta_m(\rho)&=&(\rho'+\Delta_\eta-\rho)/u_m-\alpha_q\label{mu_f},\\
\theta_m&=&(\tilde{R}_\eta/\rho')^{3/4}\frac{1}{\sqrt{m}}\label{mu_theta},\\
 u_m&=&2^{-1/3} m^{-2/3} \tilde{R}_\eta\label{mu_m},
\end{eqnarray}
\end{subequations}
where $\alpha_q$ is the $q$-root of the Airy function (i.e.\ $Ai(-\alpha_q)=0$) and $\tilde{R_\eta}=R_\eta+\Delta_\eta$ is the effective radius.
The functions $u_m$ is related to the  width and position of the mode maxima along the $\hat{r}$-axis inside the resonator. 

The previous considerations give an account for the spatial description of the field but do not reveal the spectral resonance positions. For this we use the dispersion relation \cite{Foreman:16,Demchenko:13}:

 \begin{eqnarray}
\omega_{q,p,m}^\eta\frac{n_\eta(\omega) \tilde{R}_\eta}{c}&=&l-\alpha_q\left(\frac{l}{2}\right)^{1/3}+\frac{2p(\sqrt{\tilde{R}_\eta}-\sqrt{\rho'})+\sqrt{\tilde{R}_\eta}}{2\sqrt{\rho'}}-\frac{\zeta n}{\sqrt{n^2-1}}+\frac{3\alpha^2_q}{20}\left(\frac{l}{2}\right)^{-1/3}\nonumber\\
&&-\frac{\alpha_q}{12}\left(\frac{2p(\tilde{R}_\eta^{3/2}-\rho'^{3/2})+\tilde{R}_\eta^{3/2}}{\rho'^{3/2}}\cdot \frac{2n\zeta(2\zeta^2-3n^2)}{(n^2-1)^{3/2}}\right)\left(\frac{l}{2}\right)^{-2/3}+\mathcal{O}(l^{-1}) \label{WGMeingefreq}
\end{eqnarray}
with $l=p+|m|$ and $\zeta=1$ for TE and $n^{-2}$ for TM modes.
The first part clearly relates to the standard Fabry-Perot equation and the higher order contributions take the effects of the geometry into account. 
The optical FSR between modes of the same family \{p,q\} will remain constant for several GHz but a small dispersion contribution coming mainly from the term $\propto(l^{1/3})$ and the change in the nominal refractive index $n(\omega)$ (following the Sellmeier equation \cite{zelmon_infrared_1997}).


\subsection{Second order interaction}
{To derive the rate equations describing our comb, we start from the nonlinear interaction energy which is a function of the nonlinear polarization $P^{(2)}=\chi^{(2)}E^2$ and the entire electric field $E = \sum_k E_k$. In general, $\chi^{(2)}$ is a tensor, but since all fields are parallelly polarized in our experiment, the vector equation becomes scalar. The nonlinear energy can be written as \cite{shenpaper}:}
\begin{equation}
\langle U^{(2)}\rangle = \int dV\int dt \left\langle{E}\frac{\partial {P}^{(2)}}{\partial t}\right\rangle.\label{secondenergy}
\end{equation}

The interaction Hamiltonian can be derived from this expression using the electric field operators in the second quantization which are given as: $\hat{{E}}_k(\textbf{r},t)=i\sqrt{\frac{\hbar\omega}{2\epsilon_k V_k}}(\psi_k(\textbf{r})\hat{a}_ke^{-i\omega_k t}-\psi_k^*(\textbf{r})\hat{a}_k^\dagger e^{i\omega_k t}),
$
where $\psi_k$ is the spatial mode distribution, $V_k$ is the effective mode volume, $\epsilon_k$ the permittivity, $\omega_k$ the angular frequency, and $a_k$ ($a_k^\dagger$) stand for the annihilation (creation) operator, respectively. 

First, let us consider only three interacting fields $E = E_1 + E_2 + E_3$.
The Hamilton operator for this interaction can be obtained by rewriting Eq.~(\ref{secondenergy}) using the electric field operators and assuming the energy conservation condition of $\omega_3=\omega_1+\omega_2$ (corresponding to sum frequency generation):
\begin{equation}
\hat{H}_\text{int}=2\epsilon_0 \chi^{(2)} \sqrt{\frac{\hbar^3\omega_1\omega_2\omega_3}{8\epsilon_1\epsilon_2\epsilon_3V_1V_2V_3}} \int dV (\psi_3^* \psi_2\psi_1 \hat{a}_3^\dagger\hat{a}_2\hat{a}_1 +\psi_3\psi_2^*\psi_1^* \hat{a}_3\hat{a}_2^\dagger\hat{a}_1^\dagger  ).
\end{equation}
The first term inside the integral describes the creation of a photon with frequency $\omega_3$ from the annihilation of two photons with frequencies $\omega_1$ and $\omega_2$. The second term {represents} the decay of $\omega_3$ into $\omega_1$ and $\omega_2$. Both processes in the Hamiltonian set the bidirectionally of this effect. The strength of the nonlinear interaction is governed by the overlap between the modes and the nonlinearity of the material resulting in the single photon coupling rate
\begin{equation}
g=2\epsilon_0 \chi^{(2)}  \sqrt{\frac{\hbar \omega_1\omega_2\omega_3}{8\epsilon_1\epsilon_2\epsilon_3V_1V_2V_3}} \int dV \psi_3^* \psi_2\psi_1.\label{eq:defgg}
\end{equation}
In the present system the three field distributions are rotationally symmetric (see Eq.~(\ref{WGMeq})) and therefore the integral over the angular contribution is only nonzero if the relation $m_3=m_1+ m_2$ is fulfilled \cite{reviewdmitry}. 

In a closed (lossless) system with real $\chi^{(2)}$, the nonlinear coupling constant $g$ is also real which dictates that the energy exchange between the participating fields oscillates with angular frequency $g$.
The interaction Hamiltonian becomes 
\begin{equation}
\hat{H}_\text{int}=\hbar g (\hat{a}_3^\dagger\hat{a}_2 \hat{a}_1+\hat{a}_3\hat{a}_2^\dagger \hat{a}_1^\dagger). \label{hamscat}
\end{equation}
 For the special case that one of the modes is a bright coherent field (classical field: $\hat{a}_1\rightarrow\alpha$ with $\alpha$ being the square root of the pump photon number) and no kind of depletion of this field over the relevant time scale, the Hamiltonian becomes:
\begin{equation}
\hat{H}_\text{int}=\hbar \alpha g (\hat{a}_3^\dagger\hat{a}_2+\hat{a}_3\hat{a}_2^\dagger).  
\end{equation}
This is also known as the beam splitter Hamiltonian and $\alpha g$ can be interpreted as the photon conversion rate from mode $\hat{a}_2$ to mode $\hat{a}_3$ and vice versa. The result of this kind of linearisation of the Hamiltonian is known as the zero-photon process characterized by photons scattering from one mode into another without changing the total number of photons \cite{Law}.

\subsection{$\chi^{(2)}$-comb generation}
With a strong microwave pump field the nonlinear process can cascade. In particular if multiple modes $\omega_k$ are equidistantly separated from the optical pump $\omega_0$ and have all the same nonlinear coupling constant $g$, both the sum- and difference frequency generation process cascades. Phase matching limits the interaction to only neighboring modes and the Hamiltonian can be written as:
 \begin{equation}
\hat{H}_\text{int}=\hbar |\alpha| g \sum\limits_{k=-N}^{N-1}(\hat{a}_k\hat{a}_{k+1}^\dagger +\hat{a}_k^\dagger\hat{a}_{k+1}), \label{hamsqu}
\end{equation}
where $2N+1$ is the number of modes involved in the system.
 {To describe the time evolution of this system we calculate the equations of motion of each mode's field operator using the Heisenberg picture. For a given mode $k$ it holds:
\begin{equation}
\dot{\hat{a}}_k= \frac{i}{\hbar}[\hat{H}_\text{int},\hat{a}_k]=-i\sqrt{n_\Omega}g(\hat{a}_{k-1}+\hat{a}_{k+1}),\label{eqsum}
\end{equation}
where $a_k$ denotes the $k$-th participating optical mode and $n_\Omega$ is the number of microwave photons in the system.       

Up to now we have considered an idealized system without losses. In a real system, the optical cavity is an open system which is coupled externally by a coherent microwave tone and optical laser. Moreover, there is energy loss due to material absorption, scattering, radiation loss, etc. We introduce these two features in the last equation using the electric field internal loss rate $\gamma'$ and the coupling rate $\gamma$. 
In addition, we also assume that the fields are classical, neglecting vacuum fluctuations effects. Then, we get the following equation in the rotating wave approximation for an optical pump at frequency $\omega_k'$ and microwave tone at $\Omega$:
\begin{equation}
\dot{A}_k=-(i\Delta\omega_k+\gamma+\gamma')A_k-i\sqrt{n_\Omega}g(A_{k-1}e^{i\phi_\Omega}+A_{k+1}e^{-i\phi_\Omega})+\sqrt{2\gamma}A_\text{in}\delta(\omega_k-\omega_{k'}) \label{eqtosolve}.
\end{equation}
Where $A_k$ denotes the slowly varying electric field amplitude of the mode $k$, $\Delta\omega_k=
\omega_k-(\omega_{k'}+(k-k')\cdot\Omega) $ is the total detuning between the sideband signal to its corresponding mode $k$, respectively. We have introduced the optical pump force $A_\text{in}$, which is used here to represent an optical monochromatic driving field with the amplitude given as function of the optical pump power $P_0$ as  $|A_\text{in}|=\sqrt{P_0/\hbar\omega_{k'}}$. } 
For a resonant microwave mode the microwave photon number is given by: $n_\Omega=2\gamma_\Omega P_\Omega/\hbar\Omega(\gamma_\Omega+\gamma'_\Omega)^2$, where $P_\Omega$ stands for the input power and $\gamma_\Omega$ and $\gamma'_\Omega$ stand for the coupling and internal loss field rates, respectively. 
{In steady state, the time derivatives are zero and Eq.~(\ref{eqtosolve}) becomes a simple set of linear equations:}
\begin{equation}
\begin{pmatrix}
    0\\0\\0\\\vdots\\0
\end{pmatrix}=-
\begin{bmatrix}
(1+i\nu_0)& -i\sqrt{G}e^{-i\phi_\Omega}    & i\sqrt{G} e^{+i\phi_\Omega}     &    0    &  \cdots     &0  \\
i\sqrt{G}e^{+i\phi_\Omega}  & (1+i\nu_{+1})& 0 & i\sqrt{G}e^{-i\phi_\Omega} &      \cdots   & 0 \\
\sqrt{G}e^{-i\phi_\Omega} & 0 & \ddots & \ddots & \ddots &  \vdots   \\
   0& i\sqrt{G}e^{+i\phi_\Omega}& \ddots & \ddots & \ddots & i\sqrt{G}e^{+i\phi_\Omega}  \\
 \vdots& \ddots        & \ddots & \ddots & (1+i\nu_{+N}) & 0 \\
0& 0      &    \cdots     & i\sqrt{G}e^{-i\phi_\Omega}    & 0     & (1+i\nu_{-N})
\end{bmatrix}
\begin{pmatrix}
    A_0\\A_{+1}\\A_{-1}\\\vdots\\A_{-N}
\end{pmatrix}
+\begin{pmatrix}
   \sqrt{\frac{2\gamma}{\Gamma^2}}A_\text{in}\\0\\0\\\vdots\\0
\end{pmatrix},\label{matrixfull}
\end{equation}
where we define the cooperativity $G={n_\Omega}g^2/\Gamma^2$ as a dimensionless parameter, the total coupling rate $\Gamma=\gamma+\gamma'$, and the normalized detuning of the optical sidebands $\nu_k=\Delta\omega_k/\Gamma$. 
This set of linear equations is numerically solvable for any arbitrary detuning, coupling strength or number of modes. Analytical solutions can be found for special cases and boundary conditions. We can calculate the total electric field, for an optical pump at $\omega_0$, assuming a non dispersive system (fixed free spectral range, which leads to only first order contributions in Eq.~(\ref{WGMeingefreq})), a microwave tone frequency at $\Omega=\text{FSR}$, and an infinite number of modes ($N\rightarrow\infty$): 
\begin{eqnarray}
\sum\limits_{-\infty}^\infty (1+i\nu_k)A_k+  i\sqrt{G}\sum\limits_{-\infty}^\infty(A_{k-1}e^{i\phi_\Omega}+A_{k+1}e^{-i\phi_\Omega})&=&\sum\limits_{-\infty}^\infty \sqrt{\frac{2\gamma}{\Gamma^2}}A_\text{in}\delta(\omega_k-\omega_{0})
\end{eqnarray}
We can now use the definition of the total electric field $A(t)=\sum\limits_{-\infty}^\infty A_ke^{-i\Omega k t}$ and consider only constant detuning $\nu=\nu_k=\Delta\omega/\Gamma$  and write:
\begin{eqnarray}
(1+i\nu)A(t)+  i\sqrt{G}A(t)(e^{i(\Omega t+\phi_\Omega)}+e^{-i(\Omega t+\phi_\Omega)})&=&\sqrt{\frac{2\gamma}{\Gamma^2}}A_\text{in}.\nonumber
\end{eqnarray}
The solution for the total field inside the optical system is therefore:
\begin{eqnarray}
A(t)=\frac{\sqrt{\frac{2\gamma}{\Gamma^2}}A^\text{in}}{1+i\nu+2i\sqrt{G}\cos(\Omega t+\phi_\Omega)}.\label{intrafield}
\end{eqnarray}

This solution for the field with respect to the time gives us information of the whole but not of a single sideband of the comb. From Eq.~(\ref{eqtosolve}) follows that only neighboring modes interact. Intuitively this can be understood such that a given sideband acts as pump mode for the next order sideband resulting in an exponential decay of the power of the sidebands.
We profit from this feature and use the ansatz $\sim {(e^{-\beta(G)})^k}$ for the sidebands, where $\beta(G)$ is an arbitrary function. Moreover, we can set the condition that the comb must be symmetric around the pump. Then, we can find the Fourier coefficients $A_k$ by using Eq.~(\ref{intrafield}) as follows:
\begin{eqnarray}
A(t)=\sum\limits_{-\infty}^\infty A_k(G)e^{-i\Omega k t}&=&A_0(G) \left(\sum\limits_{0}^\infty (-ie^{-\beta(G)})^ke^{-i\Omega k t-ik\phi_\Omega}  +\sum\limits_{-\infty}^{-1} (-ie^{-\beta(G)})^{-k} e^{i\Omega k t+ik\phi_\Omega}                      \right)       \nonumber\\
&=&A_0(G) \left(\sum\limits_{0}^\infty (-ie^{-\beta(G)})^ke^{-i\Omega k t-ik\phi_\Omega}  +\sum\limits_{1}^{\infty} (-ie^{-\beta(G)})^{k} e^{i\Omega k t+ik\phi_\Omega}                      \right)       \nonumber\\
&=&A_0 (G)\left(\sum\limits_{0}^\infty (-ie^{-\beta(G)})^ke^{-i\Omega k t-ik\phi_\Omega}  +\sum\limits_{0}^{\infty} (-ie^{-\beta(G)})^{k} e^{i\Omega k t+ik\phi_\Omega}   -1      \right)       \nonumber\\
&=&A_0 (G)\left(\frac{1}{1+ie^{-\beta(G)-i\Omega t-i\phi_\Omega}} +\frac{1}{1+ie^{-\beta(G)+i\Omega  t+i\phi_\Omega}}  -1      \right) \nonumber \\
&=&A_0 (G)\left(\frac{(1+e^{-2\beta} )}{1-e^{-2\beta} +2ie^{-\beta}\cos(\Omega t +\phi_\Omega)   }    \right)
\end{eqnarray}
This {can be set} equal to Eq.~(\ref{intrafield}):
\begin{eqnarray}
A_0 (G)\left(\frac{(1+e^{-2\beta} )}{1-e^{-2\beta} +2ie^{-\beta}\cos(\Omega t +\phi_\Omega)   }    \right) &=& \frac{\sqrt{\frac{\gamma}{2\Gamma^2G}}A^\text{in}}{\frac{1+i\nu}{2\sqrt{G}}+i\cos(\Omega t+\phi_\Omega)}    \nonumber\\
A_0 (G)\frac{\frac{1+e^{-2\beta}}{2e^{-\beta}}}{\frac{1-e^{-2\beta}}{2e^{-\beta}}+i\cos(\Omega t+\phi_\Omega)}&=&  \frac{\sqrt{\frac{\gamma}{2\Gamma^2G}}A^\text{in}}{\frac{1+i\nu}{2\sqrt{G}}+i\cos(\Omega t+\phi_\Omega)}.
\end{eqnarray}
%
%
%
From the last equation we find the functional dependence of $\beta$:
\begin{eqnarray}
e^{-\beta(G,\nu)}&=&\frac{1+i\nu}{2\sqrt{G}}\left(-1+\sqrt{1+4\frac{G}{(1+i\nu)^2}}\right).
\end{eqnarray}
We can use this to write down the amplitude for the pump field inside the resonator:
 \begin{eqnarray}
A_0(G,\nu)&=&\sqrt{\frac{2\gamma}{\Gamma^2}}\frac{A^\text{in}e^{-\beta}}{\sqrt{G}+\sqrt{G}e^{-2\beta}} =\sqrt{\frac{2\gamma}{\Gamma^2}}\frac{\left(-1+\sqrt{1+4\frac{G}{(1+i\nu)^2}}\right)\times(1+i\nu)A^\text{in}}{4G-(1+i\nu)^2\times\left(-1+\sqrt{1+4\frac{G}{(1+i\nu)^2}}\right)},.
\end{eqnarray}
Were the amplitudes of the sidebands decrease with their order, since $|e^{-\beta(G)}|<1$ in order for the sums above to converge. The electric field of each sideband in the cavity is given by:
\begin{eqnarray}
A_k(G)e^{-i\Omega k t}&=& \sqrt{\frac{2\gamma}{\Gamma^2}}\frac{(-ie^{-\beta(G,\nu)})^{|k|}e^{-i\Omega k t-ik\phi_\Omega} e^{-\beta(G,\nu)}}{\sqrt{G}+\sqrt{G}e^{-2\beta(G,\nu)}}A^\text{in}\label{sol2}
\end{eqnarray}
These last two equations describe quantitatively the exponential envelope for the sideband field amplitudes. Moreover, they predict that the exponential decay between neighboring sidebands decreases as $G$ increases. This leads to the OFC becoming wider with increasing parameter $G$. The limit is given by the coefficient $|e^{-\beta(G)}|\rightarrow1$ for $1/\sqrt{G}\rightarrow0$. Moreover, it follows that in the range $4G\gg1$, it holds: $\beta(G)=\frac{1}{2\sqrt{G}}$.

In a coupled system there are some external factors, such as mode overlap mismatch between the resonator and the coupler (prism), that play a role in the output power of the comb's sidebands. Furthermore, the condition of critical $(\gamma=\gamma')$, under- $(\gamma<\gamma')$ and over-coupling $(\gamma>\gamma')$ influences the output power \cite{reviewdmitry}. Therefore, the electric field amplitude of each sideband coming out from the system and the total electric field output with a perfect mode overlap is given by:
\begin{eqnarray}
A^\text{out}(t)&=&-A^\text{in}+\sqrt{2\gamma} A(t)\nonumber\\
A^\text{out}(t)=\sum\limits_{-\infty}^\infty A^\text{out}_k(t)&=&\sum\limits_{-\infty}^\infty\left(- \delta_{0,k}A^\text{in}- \frac{2\gamma}{\Gamma}A_0(G,\nu) \times(-ie^{-\beta(G,\nu)})^{|k|}e^{-i\Omega k t} \right)\nonumber\\
A^{\text{out}}(t)&=&\frac{ 2\gamma/\Gamma-1 -i(\nu +2\sqrt{G}\cos(\Omega t+\phi_\Omega))}{1+i(\nu+2\sqrt{G}\cos(\Omega t+\phi_\Omega))}A^\text{in}(t).\label{allmodes}
\end{eqnarray}
\begin{figure}
\includegraphics[width=1.0\textwidth]{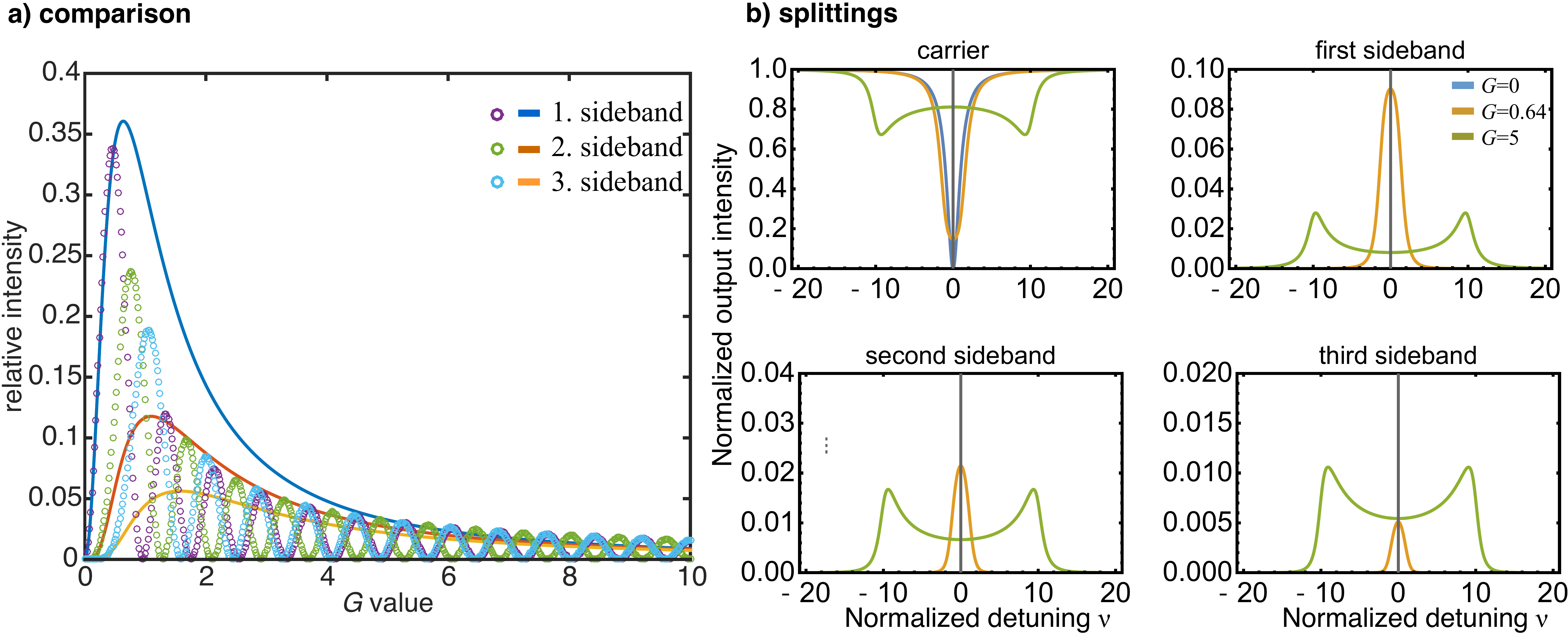} 
\caption{\label{comparsion}Comparison between the normalized sidebands of an over-coupled resonant system and a non resonant system. The sidebands of both system behave similar for low values of $G$. As $G$ increases the sidebands power of the non resonant system oscillate following the Bessel functions. On the contrary, the sidebands of the resonant system show a more stable behaviour. b) Mode splitting for a critical coupled system. The carrier and the first three sidebands splittings follow $\Delta\nu\approx4G-1.2$}
\end{figure}
This extended expression from Ref.~\cite{ilchenko_whispering-gallery-mode_2003} contains the whole information of the system such as the reflected carrier field and the sidebands' amplitude and phase. Under the condition that optic and microwave fields on are on resonance an initially strongly overcoupled configuration $\gamma/\Gamma\approx1$, maximizes the sidebands' output amplitudes:
\begin{eqnarray}
A^\text{out}(t)& \approx &\exp[-2i\arctan(2\sqrt{G}\cos(\Omega t+\phi_\Omega))]A^\text{in}(t).
\end{eqnarray}
It is worth to compare this expression with the output electric field of a non resonant phase modulator which is given by~\cite{pozar_microwave_2011}:
\begin{eqnarray}
A^\text{out}(t)&=&\exp[-i\xi\cos(\Omega t)]A^\text{in}(t)\nonumber\\
&=&A^\text{in}(t)\sum\limits_{l=-\infty}^{\infty}  (-i)^l J_l(\xi)e^{-il\Omega t}, \label{modelpm}
\end{eqnarray}
where $J_l$ are the usual $l$-th order Bessel functions of first kind and $\xi$ is the modulation index. Both expressions coincide for small values of $G$ where the small angle approximation $\arctan(x)\approx x$ holds. From this comparison, we find the {relation between} the cooperativity $G$ and the modulation index to be $\xi=4\sqrt{G}$ and the relation between the phases of the sidebands relative to the carrier \cite{hobbs2011building}.

In Fig.~\ref{fig:figure2} we show the dependence of frequency combs with respect to the cooperativity $G$, where the sidebands' power follows an exponential envelope as described in Eq.~(\ref{sol2}).  Another important aspect of this systems is given by the behaviour of the sidebands' amplitude. In a non-resonant electro-optic modulator the sidebands' power oscillates as the modulation index increases. On the other hand, in a resonant system the sidebands reach a maximum given by the initial coupling configuration and then they also decreases monotonically in $G$. {Both effects are illustrated in Fig.~\ref{comparsion}(a).}
In our system the maximum power of the first order sideband $|A^\text{out}_1|^2/|A^\text{in}_0|^2=0.36$ for $\sqrt{G}\approx 0.64$. These values depend on the number of the optical modes involved. For example, for three optical modes the first sidebands' maximal normalized power is $0.5$ at $\sqrt{g}\approx0.5$.

Another important feature of the multiple resonant system is given by the mode splitting of the reflected signal and the sidebands. This is analytically described by Eq.~(\ref{sol2}) and a critically coupled carrier behaves as:
\begin{equation}
\Delta\nu\approx 4\sqrt{G}-1.2 \mbox{ for }G>1. \label{spliform}
\end{equation}
The mode splitting of the carrier and the sidebands follow the pattern $\Delta\nu\sim4\sqrt{G}$ which is shown analytically in Fig.~\ref{comparsion}(b) and experimentally in Fig.~\ref{fig:figure2}b. The splitting in terms of absolute frequency is given by $4\sqrt{n_\Omega}g$, which is a factor two larger than in the case of two optical modes coupled with $\sqrt{n_\Omega}g$ (e.g.\ in the case for optomechanics or single sideband modulation) and a factor $\sqrt{2}$ larger than in the case of two symmetric sidebands, given as $\sqrt{2}\sqrt{n_\Omega}g$. The splitting reflects how strong the interaction between the modes are and the effective coupling strength of such a multimode system can be extracted from the eigenvalues of the dynamic matrix in Eq.~(\ref{matrixfull}). For $N\rightarrow\infty$ the matrix has infinite eigenvalues and corresponding eigenvectors. Under the constraint of a fixed $\text{FSR}=\Omega$ and driving only the central mode $A_0$, we choose the eigenvalue for the eigenvector with the highest $|A_0|$, which is given by: $\eta_\pm= \pm i\sqrt{4n_\Omega}g-2\Gamma$. This leads to the splitting $|\eta_+-\eta_-|=4\sqrt{n_\Omega}g$. The factor 1.2 in Eq.~(\ref{spliform}) changes  slightly according to the sideband order and initial coupling conditions. The splitting broadens the optical modes which then allows the comb to spread up to the sideband order whose dispersion induces a detuning $\Delta\omega_k$ which equals to $(2\sqrt{G}-0.6)\Gamma$ as shown in Fig.~\ref{fig:figure2}a. 

Another important feature intrinsically connected to the splitting is the generation of pulses. If the carrier is initially resonant and the system is critically coupled, then the reflection is ideally zero. The term $2\sqrt{G}\cos(\Omega t)$ in the denominator in Eq.~(\ref{intrafield}) acts like an effective shift of the optical resonance which oscillates with frequency $\Omega/2\pi$ and amplitude $2\sqrt{G}$. This will amplitude modulate the reflected signal with two well defined dips during a period of time $2\pi/\Omega$. The duration of the generated pulses $\Delta\tau$ is $\approx \frac{1}{\Omega \sqrt{G}}$ for the case of $4\sqrt{G}\gg1$.

From  Eq.~(\ref{sol2}) we extract information about the noise (phase or intensity) in any electrooptical modulator, which depends linearly on the sideband order. In case of the microwave phase noise $\phi_\Omega(t)$, which we assume to vary around an initial phase offset $\phi_\Omega$ with $\phi_\Omega(t)=\phi_\Omega+\Delta\phi_\Omega(t)$. The phase noise of the sideband $k$ becomes then $\sim k\Delta\phi_\Omega(t)$. This linear noise amplification of the microwave noise can be use to characterize microwave sources. On the oder hand, optical phase or intensity noise of the source is the same for each sideband order.

\subsection{Efficiency and $\pi$-Voltage}

The efficiency of an electro-optic modulator can be given in terms of V$_\pi$. This is the {required} voltage applied to an electro-optic phase modulator to {shift the phase of the optical carrier by $\pi$}. We can extract this information from the non-linear coupling constant $g$ which can be defined as the phase shift induced to the optical carrier by a single microwave photon per roundtrip time \cite{tsang_cavity_2010}:
\begin{equation}
\frac{\Delta \phi}{\tau}=g.
\end{equation}
Moreover, the classical microwave field amplitude $\alpha$ in case of a resonant microwave system is given in terms of the input power $P_\text{AC}$ as: 
\begin{equation}
n_\Omega=|\alpha|^2=\frac{2\gamma_\Omega P_\text{AC}}{\hbar\Omega((\gamma_\Omega+\gamma'_\Omega)^2+\Delta\Omega^2)},
\end{equation}
where $n_\Omega$ is the mean photon number in the mode, $\Delta\Omega$ the pump detuning, $\gamma_\Omega$ and $\gamma'_\Omega$ are the corresponding coupling and loss rates of mode fields.
The standard relation between power and a sinusoidal peak voltage $\text{V}_p$ in an AC circuit holds:
\begin{equation}
 P_\text{AC}=\text{Re}\left(\frac{\text{V}_p^2}{2Z}\right),
\end{equation}
where $Z$ is the load impedance which in most cases is \SI{50}{Ohm}. In the optical resonator the carrier stays an average time $\tau_p$ before it is outcoupled or absorbed. We find the voltage  V$_\pi$  by rewriting the relation $\alpha g \tau_p=\pi$ in terms of the voltage and it follows:
\begin{equation}
\text{V}_\pi=\frac{\pi}{ g \tau_p}\sqrt{\frac{Z\hbar\omega_\Omega(\gamma_\Omega+\gamma'_\alpha)^2 }{  \gamma_\Omega }}.
\end{equation}
For our experimental parameters, we estimate the $\pi$-voltage to be $V_\pi \approx 260\,$mV.
The ratio of the power consumption between a normal and resonant modulation is given by the ratio squared of their corresponding $V_\pi$. In our case this means that the power consumed is 100 times less than the best modulator so far reported at this frequencies. Obviously this enhancement limits the bandwidth, but in our application as a frequency comb source this does not pose a problem.

\subsection{Microwave Coupling}
We decided to use a coaxial probe coupler because it allows us to engineer a compact design for the cavity and offers an easy control over the coupling strength without modifying the rest of the cavity's geometry. The usual coupling probe is made by stripping the transmission line coaxial cable and putting the core inside the cavity leaving the outer braided shield in contact with the metal cavity. The length of the probe is ideally set to a quarter of the wavelength such that the input impedance is nearly equivalent to that of an open circuit.  The voltage difference between the probe and the adjacent metal wall of the resonator generates an electric field between them and there is also a small current flowing through the probe. The coupler's field radiates the energy similar to a monopole antenna and the electric field lines are perpendicular to the probe surface and the 3D inner surface. The pin coupler excites only modes with the electric field parallel to its own field lines and the coupling strength depends on the mode overlap of the coupler's and the cavity's modes. In this way we can choose to which mode we want to couple by locating the probe near to its maximum electric field amplitude (strong coupling) or its minimum (no coupling). We have used only one port for this experiment and we characterize the system always in reflection. A second port would enable us to also measure transmission. However, we want the photons to stay in the cavity as long as possible and once a photon is inside the cavity, a second coupling port would add another loss channel, thereby reducing the effective microwave photon lifetime.

Experimentally, a SMA connector can be fixed to the cavity's outer surface. The pin attached to it extends into the microwave cavity through a feedthrough with radius $r_h$, we choose $r_h$ such as: $0.36=\log_{10}(r_h/r_p)$ resulting in an impedance of 50\,Ohm. From the COMSOL eigenfrequency simulations we known the mode's distribution and we can place the pin close to the mode maxima to assure good mode overlap and its length can be optimized to achieve critical coupling.

\begin{figure}
\includegraphics[width=1.0\textwidth]{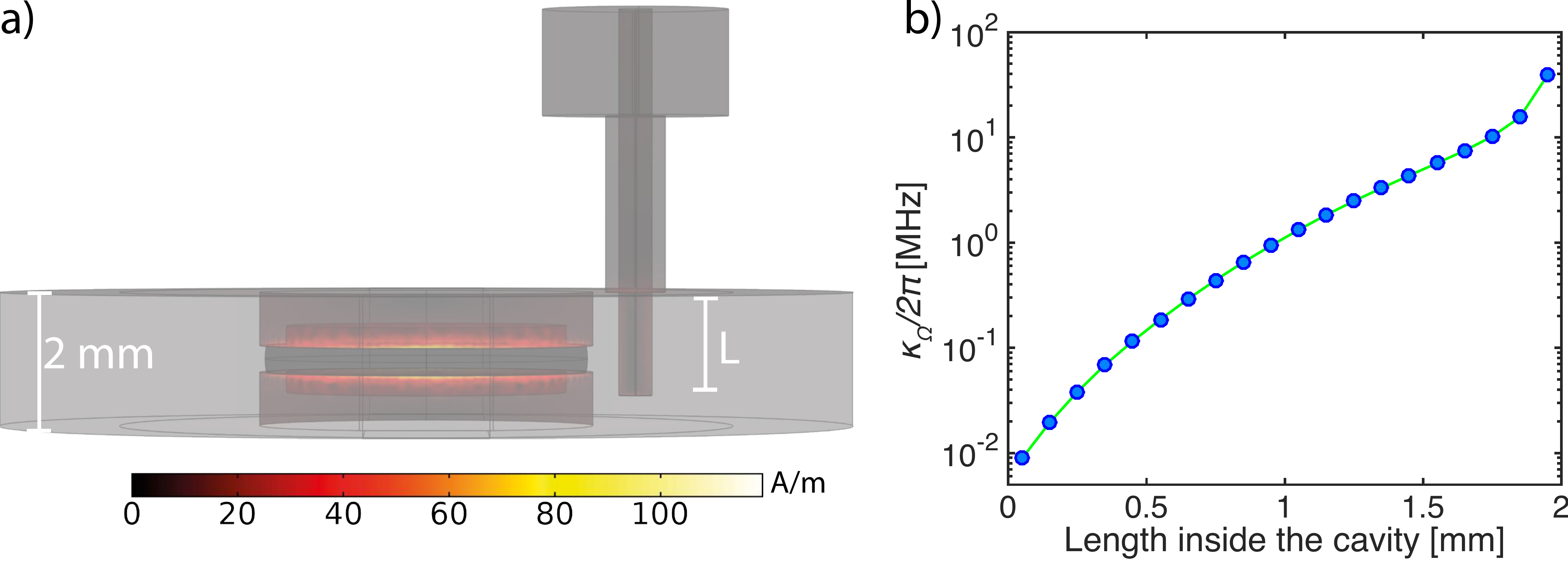}
	\caption{a) Surface current distribution inside the 3D cavity and the coaxial probe.  The length L is change from 0.05 to 1.95 mm and the excitation port power was set to 1 mW. b) Exponential behaviour of the coupling strength $\kappa_\Omega$. }
		\label{microcoupling}
\end{figure}
We ran the transient COMSOL simulations as depicted in Fig.~\ref{microcoupling}, where we have simulated  the pin's head, the feedthrough and the cavity.  The pin's head is set as coaxial port and excited over the frequencies around  the resonance given by the eigenfrequency simulation. 
The simulated coupling rate of the intensity $\kappa_\Omega=2\gamma_\Omega$ as a function of the pin's length is shown in a logarithmic plot in Fig.~\ref{microcoupling}b. The exponential growth of $\kappa_\Omega$ makes it a very sensitive parameter not only to the pin's length but also to possible bends. In our design, which has a strong electric field confinement, it requires therefore that the probe gets closer to the rings which increases the current flowing on the pin. Therefore, the pin itself contributes to ohmic losses in the system. The presence of the pin modifies also the cavity mode distribution changing the resonance frequency by several MHz.
%

\end{document}